\documentstyle[12pt]{article}
\newcommand{\Tr}{\mathop{\rm Tr}}                
\newcommand{\figurebox}[2]{\mbox{\vbox to#2in{\hbox to #1in{\hfil}
\vfil}}}

\begin{document}
\renewcommand{\thefootnote}{\fnsymbol{footnote}}
                                        \begin{titlepage}
\begin{flushright}
MPI--Ph/93--17\\
PSU/TH/126\\
April 1993\\
\end{flushright}
\vskip0.8cm
\begin{center}
{\LARGE
Instanton-induced contributions to structure functions
of deep inelastic scattering
            \\ }
\vskip1cm
{\Large Ianko I. Balitsky}~$^\ast$ \\
\vskip0.2cm
        Physics Department, Penn State University,\\
       104 Davey Lab.,University Park, PA 16802, USA\\
\vskip0.5cm
and\\
\vskip0.5cm
 {\Large Vladimir M.~Braun} $\footnote { On leave of absence from
St.Petersburg Nuclear
Physics Institute, 188350 Gatchina, Russia}$ \\
\vskip0.2cm
       Max-Planck-Institut f\"ur Physik   \\
       -- Werner-Heisenberg-Institut -- \\
        P.O.Box 40 12 12, Munich (Fed. Rep. Germany)\\
\vskip1cm
{\Large Abstract:\\}
\parbox[t]{\textwidth}{
We identify and calculate the instanton-induced contributions
to deep inelastic scattering which correspond to nonperturbative
exponential
corrections to the coefficient functions in front of parton
distributions of the leading twist.
}
\\ \vspace{1.0cm}
{\em submitted to Physics Letters B }
\end{center}
                                                \end{titlepage}

\newpage
{\bf\large 1.}\hspace{0.5cm}
The deep inelastic lepton-hadron scattering at large momentum
transfers $Q^2$ and not too small values of the Bjorken
scaling variable  $x = Q^2/2pq$ is studied in much detail and
presents a classical example for the application of perturbative
QCD. The celebrated factorization theorems allow one to separate
the $Q^2$ dependence of the structure functions in coefficient
functions $C_i (x,Q^2/\mu^2 ,\alpha_s(\mu^2))$ in front of parton
(quark and gluon) distributions of leading twist
$P_i (x,\mu^2, \alpha_s (\mu^2))$
\begin{equation}
   F_2(x,Q^2) = \Sigma_i C_i (x,Q^2/\mu^2 ,\alpha_s(\mu^2))
              \otimes
               P_i (x,\mu^2, \alpha_s (\mu^2)),
\end{equation}
where
 \begin{equation}
      C(x) \otimes P(x) = \int_x^1 \frac{dy}{y} C(x/y) P(y),
 \end{equation}
the summation goes over all species of partons, and $\mu$ is the
scale, separating "hard" and "soft" contributions to the cross
section. At $\mu^2 = Q^2$ the coefficient functions can be
calculated perturbatively and are expanded in power series in
the strong coupling
 \begin{equation}
 C(x,1,\alpha_s(Q^2)) = C_0(x)+\frac{\alpha_s(Q^2)}{\pi} C_1(x)
     +\left(\frac{\alpha_s(Q^2)}{\pi}\right)^2 C_2(x)+\ldots
 \label{cpert}
 \end{equation}
 whereas their evolution with $\mu^2$ is given by famous
 Gribov-Lipatov-Altarelli-Parisi equations.
Going over to a low normalization point $\mu^2 \sim 1 GeV$,
one obtains the structure functions expressed in terms of the
parton distributions in the nucleon at this reference scale.
The parton distributions
absorb all the information about the dynamics of large distances and
are fundamental quantities extracted from the experiment.
Provided the parton distributions are known, all the dependence
of the structure functions on the momentum transfer is calculable
and is contained in the coefficient functions $C_i$.
Corrections to this simple picture come within perturbation
theory from the parton distributions of higher twists and are
suppressed by powers of the large momentum $Q^2$.

Objective of this letter are possible exponential corrections
to the r.h.s. of eq.(\ref{cpert}) of the form
$F(x)\exp[-4\pi S(x)/\alpha_s(Q^2)]$. Since the experimental data
are becoming more and more precise, it is of acute interest to find a
boundary for a possible accuracy of the perturbative approach, set
by nonperturbative effects. Our study has also been fuelled by
recent findings of an enhancement of instanton-induced effects
at high energies
in a related problem of the violation of baryon number
in the electroweak theory \cite{ring90,matt92}, and by
an indication \cite{bal93} that in the
case of QCD the instanton-induced effects may be numerically
large at high energies,
despite the fact that they correspond formally to a contribution
of a very high fractional twist
$\exp(-4\pi S(x)/\alpha_s(Q^2)) \sim (\Lambda_{QCD}^2/Q^2)^{bS(x)}$.
Instanton-induced contributions may have a direct relation to the
contributions to structure functions of high orders of
perturbation theory \cite{zakh92,mag92}.

In this letter we consider the contribution to the structure
functions coming from the instanton-antiinstanton pair.
Contributions of single instantons are only present
for scattering from polarized targets and for higher-twist
terms in the light-cone expansion.
A Ringwald-type enhancement
\cite{ring90} of the instanton-induced cross sections at
high energy can compensate the extra semiclassical
suppression factor $\exp(-2\pi /\alpha)$ accompanying
 instanton-antiinstanton contributions
compared to single-instanton
ones.  In such case the $\bar I I$ terms become
the leading ones owing to a bigger
power of the coupling in the preexponent.

As it is well known, the instanton contributions in QCD are
in general infrared-unstable, the integrals over the instanton
size are strongly IR-divergent. Our starting point is
the observation that
this problem does not affect calculation of instanton contributions
to the coefficient functions. Let us introduce for a moment an
explicit IR cutoff $\Lambda_{IR}$ to regularize the integrals over
the instanton size. Then the contribution of
the instanton-antiinstanton pair to the cross section can be
written schematically as
\begin{equation}
     \sigma(Q^2) \sim (\Lambda_{IR}/\Lambda_{QCD})^{2b}
           + (\Lambda_{QCD}/Q)^{2bS(x)}.
\label{IR}
\end{equation}
 The second term in (\ref{IR}) gives an IR-protected contribution.
It depends in a nontrivial way on the external large momentum
and is identified  unambiguously with a contribution to
the coefficient function. The first term contributes to the
parton distribution. To be precise, one should separate in the
first term the contributions coming from instanton sizes above
and below the reference scale $\mu$, and to add
the contribution of
small-size instantons to the coefficient function.
Schematically, one has in this way
\begin{equation}
(\Lambda_{IR}/\Lambda_{QCD})^{2b} =
(\mu/\Lambda_{QCD})^{2b}  +
\left[(\Lambda_{IR}/\Lambda_{QCD})^{2b}-(\mu/\Lambda_{QCD})^{2b}
\right].
\end{equation}
However, this reshuffling of the $Q^2$-independent contribution
between the coefficient and the parton distribution does not
affect the observable cross section. It is analogous to an
ambiguity in the separation between contributions to
the coefficient function
and to the parton distribution in perturbation theory, induced by
possibility to use
different regularization schemes (e.g.
$\overline{MS}$ instead of $MS$, etc.). Hence, we can concentrate on
contributions of the second type in (\ref{IR}), which
are IR-protected.

\bigskip
{\bf\large 2.}\hspace{0.5cm}
The distinction between the instanton-induced contributions
to the coefficient functions, which are given by convergent
integrals over the instanton size, and the contributions to
parton distributions, given in general by IR-divergent integrals,
becomes especially transparent for an example of the
cross section of hard gluon scattering from a real gluon, see Fig.1a,
 considered
in detail in \cite{bal93}. In this case one needs to evaluate the
contribution to the functional integral
coming from the vicinity of the instanton-antiinstanton
configuration (and amputate external gluon legs afterwards).
Each hard gluon is substituted by the Fourier
transform of the instanton field at large momentum, and brings in
the factor
\cite{andrei}
\begin{equation}
   A_{\nu}^I(q)  \simeq
    \frac{i}{\rm g} (\sigma_{\nu}\bar{q} - q_{\nu})
    \left\{\frac{8\pi^2}{Q^4}-(2\pi)^{5/2}
\frac{\rho^2}{2Q^2}(\rho  Q)^{-1/2} e^{-\rho Q} .
\right\}
\label{1i}
\end{equation}
The first term in (\ref{1i}) produces a power-like divergent
integral over the instanton size $\rho$. All dependence on the
hard scale comes in this case through the explicit power
of $Q^2$ in front of the divergent integral. This is  a typical
contribution to the parton distribution --- in the present case
to the probability to find a hard gluon within a soft gluon.
The second term gives rise to a completely different behavior.
The cross section is given in this case by the following
integral over the common scale of the instanton and antiinstanton
$\rho_I \sim \rho_{\bar I}$ and over their separation $R$ in the c.m.
frame:
\begin{equation}
\int \rho\,d\rho\int dR_0\,\exp\left\{-2Q\rho+ER_0-
\frac{4\pi}{\alpha_s(\rho)}S(\xi)\right\}.
\label{1q}
\end{equation}
Three important ingredients in this expression are: the factor
$\exp(-2Q\rho)$, which comes from the two hard gluon fields,
the factor $\exp(ER_0)$, which is obtained from the standard
exponential factor $\exp(i(p+q)R), E^2 = (p+q)^2$ by the rotation
to Minkowski space, cf. \cite{zakh90}, and the action
 $S(\xi)$  evaluated on the instanton-antiinstanton
configuration. The normalization is such that $S(\xi)=1$ for
an infinitely separated instanton and antiinstanton, and $\xi$
is the conformal parameter \cite{yung88}
\begin{equation}
  \xi=\frac{R^2+\rho_1^2+\rho_2^2}{\rho_1\rho_2}.
\label{xi}
\end{equation}
Writing the action as a function of $\xi$ ensures that the
interaction  between instantons is small in two different limits:
for a widely separated $I \bar I $ pair, and for a small instanton
put inside a big (anti)instanton, which are related to each other
by the conformal transformation. In the limit of large $\xi$ the
expansion of $S(\xi)$ for the dominating
maximum attractive $I \bar I $ orientation reads \cite{yung88}
\begin{equation}
  S(\xi) =  \left(1-\frac{6}{\xi^2}+O(\ln(\xi)/\xi^4)\right)
\label{action}
\end{equation}
where the $1/\xi^2$ term corresponds to a slightly corrected
 dipole-dipole
 interaction. In general, one can have in mind a certain
smooth function which turns to zero at $R\rightarrow 0$.

 To the semiclassical accuracy the integral in (\ref{1q})
is evaluated by a saddle-point method.
 The saddle-point equations take the form \cite{bal93}
\begin{eqnarray}
 Q\rho_\ast &=&
\frac{4\pi}{\alpha_s(\rho_\ast)}(\xi_\ast-2)S'(\xi_\ast)
+bS(\xi_\ast)\,,
        \nonumber\\
E \rho_\ast=  &=&\frac{8\pi}{\alpha_s( \rho_\ast)}
\sqrt{\xi_\ast-2}\,S'(\xi_\ast)\,,
\label{saddleeq}
\end{eqnarray}
where $S'(\xi)$ is the derivative of $S(\xi)$ over $\xi$, and
$\rho_\ast, \xi_\ast$ are the saddle-point values for the
instanton size and the conformal parameter, respectively.

Neglecting in (\ref{saddleeq}) the terms proportional to
$b=(11/3) N_c - (2/3)n_f$, which come from the differentiation
of the running coupling and only produce a small correction,
one finds
\begin{eqnarray}
      \xi_\ast & = & 2 + \frac{R_\ast^2}{\rho^2_\ast}
        = 2 \frac {1+x}{1-x} ,
\nonumber\\
     Q\rho_\ast &=& \frac{4\pi}{\alpha_s(\rho_\ast)}
     \frac{12}{\xi_\ast^2}
\label{saddle}
\end{eqnarray}
A numerical solution of the saddle-point equations in (\ref{saddleeq})
for the particular expression of the action $S(\xi)$ corresponding
to the conformal instanton-antiinstanton
valley is shown in Fig.2.\footnote{
Numerical results are strongly sensitive to the particular value
of the QCD scale. We use the two-loop expression for the coupling
with three active flavors,
and the value $\Lambda_{\overline MS}^{(3)} = 290 MeV$ which
corresponds to  $\Lambda_{\overline MS}^{(4)} = 240 MeV$ \cite{PDT}.
The corresponding value of the coupling is
$\alpha_s (1 GeV) = 0.41$.
}
Note that the difference between the hard scale $Q^2$ and the
effective scale for nonperturbative effects $\rho^{-2}_\ast$ is
numerically very large.  This is a new situation compared to
 calculations of instanton-induced
 contributions to two-point correlation functions, see
e.g. \cite{andrei,NSVZ80,DS80}, where the
size of the instanton is of order of the large virtuality, and
indicates that the instanton-induced contributions to deep
inelastic scattering may turn out to be nonnegligible even at
values
$Q^2 \sim 1000 GeV^2$, which are conventionally
considered as a safe domain for perturbative QCD.

In the case of hard gluon-gluon scattering it is easy to collect
all the preexponential factors (to the semiclassical accuracy).
The result for the scattering of a transversely polarized hard
gluon from a soft gluon reads \cite{bal93}\footnote{
In difference to \cite{bal93} we give the result for scattering
of gluons with fixed color and polarization.}
\begin{eqnarray}
2E^2\sigma_{\perp} & =&
\frac{4}{9} d^2 \frac{(1-x)^2+x^2}{x^2 (1-x)^2}
\pi^{13/2} \left(\frac{2\pi}{\alpha(\rho_\ast)}\right)^{21/2}
\exp \left[ -
\left(\frac{4\pi}{\alpha(\rho_\ast)}+2b\right)
S(\xi_\ast )\right]\,.
\label{gluon}
\end{eqnarray}
 It is expressed in terms of the saddle-point values of
  $\rho$ and $\xi$.
Here $d\simeq 0.00363$ (for $n_f=3$) is a constant which enters
the expression for the instanton density
\begin{equation}
 d  =  \frac{1}{2}C_1 \exp[n_f C_3-N_c C_2)],
\label{d}
\end{equation}
$C_1 = 0.466, C_2 = 1.54, C_3 = 0.153$ in the $\overline{MS}$
scheme.
At $\alpha_s(\rho_\ast) \simeq 0.3-0.4$ and  $S(\xi_\ast)\simeq 1/2$
the expression on the r.h.s. of (\ref{gluon}) is of order
$10^{-2}-10^{0}$,
which means that at $Q^2 \sim 100-1000 GeV^2$ and $x< 0.25-0.40$
the nonperturbative contribution is significant.

\bigskip
{\bf\large 3.}\hspace{0.5cm}
Similar contributions are present  in the structure functions
of deep inelastic lepton-hadron scattering, but the calculation
turns out to be much more involved. The situation appears to be
simpler for the case of deep inelastic scattering from a real
gluon. To this end we need to evaluate
\begin{eqnarray}
T_{\mu\nu}& =& i\int d^x\,e^{iqx} \langle A^a(p),\lambda |
T\{ j_\mu(x) j_\nu (0)\}|A^a(p),\lambda\rangle
\nonumber\\
W_{\mu\nu}&=&\frac{1}{\pi} \mbox{Im}\, T_{\mu\nu}=
         \\
&&\mbox{} =
\left(-g_{\mu\nu}+\frac{q_\mu q_\nu}{q^2}\right)F_L(x,Q^2)
+\left(\frac{p_\mu p_\nu}{pq}-
\frac{p_\mu q_\nu + q_\mu p_\nu}{q^2} +g_{\mu\nu}
\frac{pq}{q^2}\right) F_2(x,Q^2)
\nonumber
\label{str}
\end{eqnarray}
We have found it simpler to do the calculations in the
coordinate space. Full details will be published elsewhere
\cite{bal94}, and below we only indicate the main steps.

 The true small parameter in our calculation is the value of the
coupling constant at the scale $Q^2$, which ensures that the effective
instanton size is sufficiently small.
We are trying to be accurate to collect to the semiclassical
accuracy all the dependence on $\rho^2/R^2$ in the exponent,
having in mind the valley method \cite{bal86}, in which
all the dependence on the $\bar I I$
separation is absorbed in the action $S(\xi)$ on the $\bar I I$
configuration. However, in this letter we do not take into
account corrections of order $\rho^2/R^2$ in the preexponent,
and to this accuracy
need the first nontrivial term only in the
cluster expansion of the quark propagator at the $\bar I I$
background \cite{andrei}:
 $
  \langle x|\nabla_{I\bar I}^{-2}\bar \nabla_{I\bar I}|0\rangle
 =
 \int dz\, \langle x|\nabla_1^{-2} \bar\nabla_1|z\rangle
 \sigma_\xi \frac{\partial}{\partial z_\xi}\langle z|
 \bar\nabla_2 \nabla_2^{-2} |0\rangle
 $.

 The leading contribution to the gluon matrix
element of the T-product of the electromagnetic currents in
(\ref{str}) is given by the following expression
\begin{eqnarray}
\lefteqn{
\langle A^a(p),\lambda |T\{ j_\mu(x) j_\nu (0)\}|A^a(p),\lambda\rangle
^{I\bar{I}}         =}
\nonumber\\
 &&\mbox{} = \sum_q e^2_q
 \int dU
\int \frac{d\rho_1}{\rho_1^5} d(\rho_1)
\int \frac{d\rho_2}{\rho_2^5} d(\rho_2)
\int d^4 R \int d^4 T
\nonumber\\
&&\mbox{}\times
\frac{1}{8}
\,\,\mbox{lim}_{p^2\rightarrow 0}\, p^4
\epsilon^\lambda_\alpha \epsilon^\lambda_\beta
\Tr\left\{A_{\alpha}^{\bar{I}}(p) A_{\beta}^I(-p)\right\}
\exp\left[-\frac{16\pi^2}{{\rm g}^2} S^{\bar I I}\right]
\nonumber\\
 &&\mbox{}\times
 (a^\dagger a)^{n_f-1}
\left\{ a  \bar \phi_0(0)\bar\sigma_\nu
\langle 0|\nabla_2^{-2}\nabla_2\bar\partial \nabla_1 \nabla_1^{-2}
|x\rangle \bar\sigma_\mu \phi_0(x)
\right.
 \nonumber\\
 &&\mbox{} \left.
 +a^\dagger
\bar \kappa_0(x) \sigma_\mu
 \langle x|\nabla_1^{-2} \bar
 \nabla_1\partial \bar\nabla_2 \nabla_2^{-2}
|0\rangle \sigma_\nu \kappa_0(0)
+(\mu \leftrightarrow \nu, x \leftrightarrow 0)  +\ldots
\right\}
\label{formula1}
\end{eqnarray}
which corresponds to the diagram shown in Fig.1b.
The full expression contains many more terms \cite{bal94}
which are not shown because we have found that all of them are
of order $O(\alpha_s(Q^2))$ compared to the expression in
(\ref{formula1}).
Here and below the subscript '1' refers to the antiinstanton
with the size $\rho_1$ and the position of the center
 $x_{\bar I} = R+T$,
and the subscript '2'  refers to the instanton with the size
$\rho_2$ and the center at $x_I = T$.
 We use   conventional notations $\nabla = \nabla_\mu\sigma_\mu$ and
$\bar\nabla = \nabla_\mu\bar\sigma_\mu$, etc,  where
$\sigma_\mu^{\alpha\dot\alpha} = (-i\sigma, 1)$,
$\bar\sigma_{\mu\dot\alpha\alpha} = (+i\sigma, 1)$, and
$\sigma$ are the standard  Pauli matrices.
The expressions for quark propagators at the
one-instanton (antiinstanton)
background are given in  \cite{brown}.

  The quark zero modes  are written in terms of
two-component Weil spinors
 $\psi_0 = \left(
 \begin{array}{c} \kappa_0\\ \phi_0 \end{array}\right)$,
$\psi^\dagger_0 =
\left( \bar \phi_0\,\, \bar\kappa_0 \right)$,
explicit expressions for which in case of
 the instanton (antiinstanton) with the center at the
origin are
\begin{eqnarray}
\kappa^k_{\dot\alpha}(x) =
\epsilon^{\alpha\beta} (U_I)^k_\beta
 \frac{\bar x_{\dot\alpha\alpha}}{2\pi^2x^4}
\frac{2\pi \rho_2^{3/2}}{\Pi_{2x}^{3/2}}
&,&
 \bar \phi^{\dot\alpha}_k (x) =
 \epsilon_{\beta\alpha}(U^\dagger_I)^\beta_k
 \frac{x^{\alpha\dot\alpha}}{2\pi^2x^4}
\frac{2\pi \rho_2^{3/2}}{\Pi_{2x}^{3/2}}
\nonumber\\
\phi^{k\alpha}(x) =
\epsilon_{\dot\alpha\dot\beta}(U_{\bar I}u_0)^{k\dot\beta}
 \frac{x^{\alpha\dot\alpha}}{2\pi^2x^4}
 \frac{2\pi \rho_1^{3/2}}{\Pi_{1x}^{3/2}}
 &,&
 \bar\kappa_{k\alpha}(x) =
\epsilon_{\dot\alpha\dot\beta}(\bar u_0
U_{\bar I}^\dagger)_{\dot\beta k}
 \frac{\bar x_{\dot\alpha\alpha}}{2\pi^2x^4}
 \frac{2\pi \rho_1^{3/2}}{\Pi_{1x}^{3/2}}
\label{zero}
\end{eqnarray}
where $\Pi_{1x}= 1+\rho_1^2/x^2$ and $\Pi_{2x}= 1+\rho_2^2/x^2$,
respectively.
The corresponding overlap integrals are equal to
\begin{eqnarray}
   a = - \int\!dx\, (\bar\kappa \partial \kappa) =
  2\xi^{-3/2} \mbox{Tr}\,O
&,&
 a^\dagger  = - \int\!dx\,(\bar\phi\bar\partial\phi)
  = 2\xi^{-3/2} \mbox{Tr}\,O^\dagger
\end{eqnarray}
where $O$ is the left upper $2\times2$ corner of the matrix
$U^\dagger_I U_{\bar I}u_0$ of the relative $I\bar I $ orientation.
The integration over the $\bar I I$ orientations in (\ref{formula1})
can be done in the
saddle-point approximation.
Assuming the standard orientation dependence of the action
corresponding to the dipole-dipole interaction, we get
\cite{bal93} (for $N_c=3$):
\begin{equation}
\int dU\, e^{-\frac{16\pi^2}{{\rm g}^2} S(\xi,U)}=
\frac{1}{9\sqrt{\pi}}
\left(\frac{\alpha_s}{2\pi}\frac{\xi^2}{4}\right)^{7/2}
 e^{-\frac{16\pi^2}{{\rm g}^2} S(\xi)} .
\label{1p}
\end{equation}
The saddle point is achieved at the maximum attractive
$I\bar I$ orientation  corresponding to the choice
$U^\dagger_I U_{\bar I}=1$,
$  u_0 = R/\sqrt{R^2}$.
The factor coming from soft gluons is
\begin{eqnarray}
\frac{1}{8}
\,\,\mbox{lim}_{p^2\rightarrow 0}\, p^4
\epsilon^\lambda_\alpha \epsilon^\lambda_\beta
\Tr\left\{A_{\alpha}^{\bar{I}}(p) A_{\beta}^I(-p)\right\}
&=& -\frac{\pi^4}{{\rm g}^2} \rho_1^2\rho_2^2
     e^{ipR}
\epsilon^\lambda_\alpha \epsilon^\lambda_\beta
 \Tr\left\{
\sigma_{\beta}\bar{p}
u_0\bar\sigma_{\alpha}p\bar{u_0}\right\}
\nonumber\\
&&\mbox{} =
\frac{\pi^3}{\alpha_s}(pR)^2 \frac{\rho_1^2\rho_2^2}{R^2} e^{ipR}
\end{eqnarray}
Main complication comes from the quark propagator. Using explicit
expression for the zero modes in (\ref{zero}) and the propagators
from \cite{brown} we find
\begin{eqnarray}
\lefteqn{\bar \kappa_0(x) \sigma_\mu
 \langle x|\nabla_1^{-2} \bar
 \nabla_1\partial \bar\nabla_2 \nabla_2^{-2}
|0\rangle \sigma_\nu \kappa_0(0) =}
\\
&&\mbox{} =
-\frac{1}{4\pi^6} \int dz \frac{(\rho_1\rho_2)^{3/2}}
{\Pi_{1x}^2\Pi_{20}^2 (x-R-T)^4 T^4}
\frac{1}{\sqrt{R^2}} \mbox{Tr}
\left\{
\frac{\bar\sigma_\nu z \bar\sigma_\xi (x-z)\bar\sigma_\mu}
     {(x-z)^4 z^4}
\right.
 \nonumber\\
 &&\mbox{} \times
\left.
\left[(x-R-T)+\rho_1^2\frac{(z-R-T)}{(z-R-T)^2}\right]
\bar R \frac{1}{\sqrt{\Pi_{1z}}}\frac{\partial}{\partial z_\xi}
 \frac{1}{\sqrt{\Pi_{2z}}}
 \left[T-\rho_2^2\frac{(z-T)}{(z-T)^2}\right]
\right\} +\ldots
\nonumber
\label{prop1}
\end{eqnarray}
Omitted terms have turned out to be of order $O(\alpha_s)$.

We have found that in order to pick up the leading contribution
in the strong coupling
one can have in mind the following mnemonic rules for the
essential regions of integration:
\begin{eqnarray}
      z^2 &\sim & (z-x)^2  \sim  x^2
\nonumber\\
     (x-R-T)^2+\rho_1^2&\sim & T^2+\rho_2^2  \sim  x^2
 \nonumber\\
     (z-R-T)^2+\rho_1^2 &\sim &  (z-T)^2 +\rho_2^2\sim x^2/\alpha_s
\nonumber\\
     T^2 &\sim & R^2 \sim \rho_1^2\sim \rho_1^2 \sim  x^2/\alpha_s
\end{eqnarray}
Note that all the calculation is done in Euclidian space, and the
evaluation of integrals by means of the analytical continuation
effectively corresponds to going over to negative values of $\rho^2$.

Hence the integration over $z$ in (\ref{prop1}) can be done in
the "light-cone" approximation:
\begin{equation}
\int dz\, \frac{F(z)}{(x-z)^4 z^4} =
\frac{\pi^2}{x^4} \int^1_0
d\gamma \gamma^{-1} \bar \gamma^{-1}
 F(\gamma x) +O(\sqrt{\alpha_s}),
\end{equation}
where $F(z)$ is an arbitrary function containing all other
possible denominators like $(z-R-T)^2+\rho_1^2$ etc.
Here and below we use the notation
$$\bar\gamma = 1-\gamma .$$
etc.
To this accuracy a simpler expression for (\ref{prop1}) can be written:
\begin{eqnarray}
\lefteqn{
-\frac{1}{2\pi^4x^4} \int_0^1 d\gamma \frac{(\rho_1\rho_2)^{3/2}}
{[(x-R-T)^2+\rho_1^2]^2[T^2+\rho_2^2]^2 \sqrt{R^2}}    }
\\
&&\mbox{}\times
\mbox{Tr}\left\{
\bar\sigma_\nu x \bar\sigma_\mu
\left[ \bar\gamma x \frac{1}{\sqrt{\Pi_{1\gamma}}}
+(\gamma x -R-T)\sqrt{\Pi_{1\gamma}}\right]
\bar R \frac {d}{d\gamma}
 \left[\gamma x\frac{1}{\sqrt{\Pi_{2\gamma}}}
 -(\gamma x -T)\sqrt{\Pi_{2\gamma}}\right]
\right\}
\nonumber
\end{eqnarray}
where $\Pi_{1\gamma}=1+\rho^2_1/(\gamma x-R-T)^2 $ and
 $\Pi_{2\gamma}=1+\rho^2_2/(\gamma x-T)^2 $.
To do the remaining integrals, we expand the exponential factor
corresponding to the dipole-dipole $I \bar I$ interaction
\begin{equation}
\exp\left [\frac{24\pi}
{\alpha_s}\frac{\rho_1^2\rho_2^2}{R^4}\right ]
=\sum_{n=0}^\infty
\frac{1}{n!} (6b)^n \left( \frac{\rho_1^2\rho_2^2}{R^4}\right)^n
\left(\ln\frac{1}{\rho_1\rho_2\Lambda^2}\right)^n .
\label{expand1}
\end{equation}
The logarithms come from the running of the coupling.
To take them into account, one needs to replace
 $n\rightarrow n+\epsilon$, do all the integrations, and
 pick up
the n-th term of the expansion
in $\epsilon$ at the end.
At the cost of this expansion
the dependence on $\rho_1$ and on $\rho_2$  factorize from each other.
The corresponding integrals are divergent and must be
understood in the sense of analytical
continuation. For definiteness, one can have in mind the following
representation
\begin{eqnarray}
\int_0^\infty d\rho^2 \frac{(\rho^2)^{\mu+n-1} \Gamma(\lambda)}
{(T^2+\rho^2)^\lambda} & = &
\frac{\Gamma(\lambda)}{2i \sin [\pi (\lambda-\mu-n)]}
\int_{-\infty}^0\!d\rho^2\,(-\rho^2)^{\mu+n- \lambda -1}
\nonumber\\
&&\mbox{}
\times
\left[\left(\frac{\rho^2+ i\epsilon}{T^2+\rho^2+ i\epsilon}
\right)^\lambda -\mbox{c.c.}\right]
 = \frac{\Gamma(\lambda-\mu-n)\Gamma(\mu+n)}{(T^2)^{\lambda-\mu-n}}
\label{rhoint}
\end{eqnarray}
The master formula for doing the remaining integrations reads
\begin{eqnarray}
\lefteqn{
\sum_{n=0}^{\infty} \left(\frac{24\pi}{\alpha_s}\right)^n
\frac{1}{n!}
\int\!dT
\int\frac{d\rho_1^2}{\rho_1^2}(\rho_1^2)^{\mu_1+n}
\int\frac{d\rho_2^2}{\rho_2^2}(\rho_2^2)^{\mu_2+n}
\int\!dR \frac{e^{ipR}}{R^{4n+2}}
\frac{\Gamma(\lambda_1)}{[(x-R-T)^2 +\rho_1^2]^{\lambda_1} }  }
\nonumber\\
&&\mbox{}\times
 \frac{\Gamma(\lambda_2)}{[T^2 +\rho_2^2]^{\lambda_2} }
\frac{[(\gamma x-R-T)^2 ]^{-\sigma_1}}
{[(\gamma x-R-T)^2 +\rho_1^2]^{\nu_1} }
 \frac{[(\gamma x-T)^2 ]^{-\sigma_2}}
 {[(\gamma x-T)^2 +\rho_2^2]^{\nu_2} }
 N(x,R,T)
\nonumber\\
&=&
\!\pi^3 \!\int_0^1\!\!du\, (2 u)^{\theta-3}
\exp\!\left[ipxu +\frac{3\pi}{2\alpha_s}
\frac{\bar u^2}{u^2} \right]
\!\left(\frac{3\pi}{2\alpha_s}\frac{\bar u^2}{u^2}\right)^{
\mu_1+\mu_2-\nu_1-\nu_2-\sigma_1-\sigma_2-2}
\!\!\left(\frac{\bar u}{2\bar\gamma}\right)^{\nu_1}
\!\!\left(\frac{\bar u}{2\bar\gamma-\bar u}\right)^{\sigma_1}
\nonumber\\
&&\mbox{}
\times
\left(\frac{\bar u}{2\gamma}\right)^{\nu_2}
\! \left(\frac{\bar u}{2\gamma-\bar u}\right)^{\sigma_2}
\int\!dR\,e^{ipR}
\frac{\Gamma(\theta-1)}{[R^2+ \bar u u x^2]^{\theta-1}}
N(x,R+ux,\frac{1}{2}(\bar u x -R)) \,,
\end{eqnarray}
where $\theta = \nu_1+\nu_2+\sigma_1+\sigma_2+\lambda_1+\lambda_2
-\mu_1-\mu_2$ and the function $N(x,R,T)$ collects the factors in the
numerator. The final integration over $R$ is elementary.

After considerable algebra we obtain the following result,
where we have added the symmetric to (\ref{formula1}) contribution
with the instanton replaced by the antiinstanton:
\begin{eqnarray}
\lefteqn{
\langle A^a(p),\lambda |T\{ j_\mu(x) j_\nu (0)\}|A^a(p),\lambda\rangle
^{I\bar{I}}         =}
\nonumber\\
 &&\mbox{} =
i \left (x_\mu p_\nu + x_\nu p_\mu -\delta_{\mu\nu} (px) \right)
 \sum_q e^2_q  \frac{2}{9} \pi^{5/2} d^2
\left(\frac{2\pi}{\alpha_s}\right)^{19/2}
\nonumber\\
&&\mbox{}\times
\int_0^1\!du\, \cos(upx) \frac{u}{\bar u^2}
\left(\frac{16}{\xi^3}\right)^{n_f-3}
\frac{\Gamma [-b S(\xi)]}{x^4} J(u)
\exp\left[ -\frac{4\pi}{\alpha_s(\tilde \rho^2)}S(\xi)\right]
\label{canswer}
\end{eqnarray}
The effective scale of the coupling under the exponential
is equal to
\begin{equation}
\tilde \rho^2 = x^2 \frac{\bar u}{2} \frac{6}{\xi^2}
\frac{4\pi}{\alpha_s(\tilde \rho^2)}
\label{xscale}
\end{equation}
and  the function $J(u)$ is defined as
\begin{eqnarray}
J(u) &=& 4 \int \!d\gamma\,
\mbox{Re}\left\{
\frac{1-i\cot[\pi b S(\xi)/2]}
{\sqrt{1-\bar u/(2 \bar\gamma) +i\epsilon}}
\right\}
\frac{d}{d\gamma}
\mbox{Re}\left\{
\frac{1-i\cot[\pi b S(\xi)/2]}
{\sqrt{1-\bar u/(2 \gamma) +i\epsilon}  }
\right\}
\nonumber\\
&&\mbox{}
= 1  +O(\bar u)
\end{eqnarray}
To the accuracy of our approximation, which was in taking
into account
the dipole-dipole term only in the
 $\bar I I $ interaction,  we obtain
$\xi = 4u/\bar u $, and $ S(\xi) = 1-6/\xi^2$.
It is easy to trace that
writing the interaction in the
conformal form (\ref{xi}),(\ref{action}) amounts to
the substitution
$\xi \rightarrow 2+4u/\bar u = 2(1+u)/(1-u)$, cf.(\ref{saddle}).
The integrand in (\ref{canswer})
is  exact for
 $1-u \ll 1$ and to leading accuracy in $\alpha_s(x^2)$.

Making a Fourier transformation, going over to the Minkowski
space and taking the imaginary part (\ref{str}), we obtain the
following answer for the instanton-antiinstanton contribution
to the structure function of a real gluon:
\begin{eqnarray}
F_1^{(G)}(x,Q^2) &=&\sum_q e^2_q
\frac{1}{9\bar x^2}
\frac{ d^2\pi^{9/2}}{bS(\xi_\ast)[bS(\xi_\ast)-1]}
\left(\frac{16}{\xi_\ast^3}\right)^{n_f-3}
\nonumber\\
&&\mbox{}\times
\left(\frac{2\pi}{\alpha_s(\rho_\ast^2)}\right)^{19/2}
\!\!\exp\left[ -
\left(
\frac{4\pi}{\alpha_s(\rho_\ast^2)} +2b\right)
S(\xi_\ast)\right]
\label{answer}
\end{eqnarray}
where the expressions for $\rho_\ast$ and $\xi_\ast$ coincide
to the ones given in (\ref{saddle}), which correspond to the direct
evaluation of the integrals in the momentum space by the saddle-point
method. Encouraged by this coincidence, we have put by hands
an additional factor $\exp(-2bS(\xi_\ast))$ to the r.h.s. of
(\ref{answer}), which arises from taking into account
the running of the
 coupling  in the saddle-point equations in (\ref{saddleeq}) and
which is difficult to trace starting from the coordinate space.
To our accuracy, we find that the instanton-
induced contributions obey the Callan-Gross relation
$F_2(x,Q^2)= 2x F_1(x,Q^2)$.

The expression in (\ref{answer}) presents main result of this letter.
It gives  the exponential correction to the coefficient
function in front of the gluon distribution of the leading twist
in (\ref{cpert}).
The exponential factor is exact to the accuracy of (\ref{action}).
Taking into account further terms in the expansion of the action
in inverse powers of the conformal parameter, one should in principle
add the contributions of rescattering in the initial state
(the so-called hard-hard corrections \cite{MU91}).
The preexponential factor in (\ref{answer})
is calculated to leading accuracy
in the strong coupling and up to corrections of order $O(1-x)$.
 The corresponding to (\ref{answer})
contribution to the structure function of the nucleon is obtained
in a usual way, making a convolution of (\ref{answer}) with a
distribution of gluons in the proton at the scale $\rho_\ast^2$.
Leading contribution to (\ref{answer}) in the perturbation theory
is due to the mixing with the flavor-singlet quark distribution
and is given by the box graph:
\begin{eqnarray}
F_1^{(G)}(x,Q^2)_{pert} =  \sum_q e^2_q
[x^2+\bar x^2]\frac{\alpha_s(Q^2)}{2\pi}
\ln\left[\frac{Q^2\bar x}{\mu^2 x}\right] ,
\label{box}
\end{eqnarray}
where in order to compare to the instanton
contribution in (\ref{answer}) one should choose the scale $\mu$
to be of order $\rho_\ast$.

The instanton-antiinstanton contribution to the structure function
of a quark contains a similar contribution shown in Fig.1c.
The difference to the graph in Fig.2b considered above is in
trivial factors only, and the answer for this contribution reads
\begin{eqnarray}
F_1^{(q)}(x,Q^2) &=&
\left[\sum_{q'\neq q} e^2_{q'} +\frac{1}{2}e^2_q\right]
 \frac{128}{81\bar x^3}
\frac{d^2\pi^{9/2} }{bS(\xi_\ast)[bS(\xi_\ast)-1]}
\left(\frac{16}{\xi_\ast^3}\right)^{n_f-3}
\nonumber\\
&& \mbox{}\times
\left(\frac{2\pi}{\alpha_s(\rho_\ast^2)}\right)^{15/2}
\exp\left[ -
\left(
\frac{4\pi}{\alpha_s(\rho_\ast^2)} +2b\right)
S(\xi_\ast)\right]
\label{qanswer}
\end{eqnarray}
However, in this case additional contributions exist of the
type shown in Fig.1d. They are finite (the integral over
instanton size is cut off at $\rho^2 \sim x^2/\alpha_s$
), but the instanton-antiinstanton separation
 $R$ does not stay large at sufficiently small energies (large
values of x), but always appears to be of order $\rho$.
This probably means that the structure of nonperturbative
contributions to quark distributions is more complicated.
This question is under study.
The answer given in (\ref{qanswer}) presents the
contribution of the particular saddle point in (\ref{saddle}).

\bigskip
{\bf\large 4.}\hspace{0.5cm}
 The region of validity of the above expressions for the
$\bar I I$ contributions to structure functions is
restricted by the requirement that the effective value
of the instanton size is not too large, say $\rho_\ast< 1 GeV^{-1}$.
 This sets a boundary
for the lowest possible value of $Q^2$ of order $100 GeV^2$, see
Fig.2.
The instanton-induced contribution to the structure function of a
gluon in (\ref{answer}) is shown as a function of Bjorken x for
different values of $Q\sim 10-100 GeV$ in Fig.3. The contribution of
the box graph in (\ref{box}) is plotted by dots for comparison.
It is seen that the $\bar I I$ contribution is rising very rapidly
with the decrease of x, the effect being due mainly to the
decrease of the $\bar I I$ action $S(\xi_\ast)$.
We take $S(\xi_\ast)=1/2$ as a reasonable boundary for which
our calculation is justified, which
translates to the condition that $x>0.3-0.35$.
We expect that at
$S(\xi_\ast)\simeq 0.5$ the accuracy of  (\ref{answer})
 is  within one order of magnitude.
Thus,  instantons produce a well-defined and calculable
contribution to the cross section of deep inelastic scattering
 for sufficiently large values of x and large $Q^2\sim 100 - 1000
GeV^2$,
which turns out, however, to be rather small ---
of order $10^{-2}-10^{-5}$ compared to the
perturbative cross section.
This  means that the
accuracy of standard perturbative analysis is sufficiently
high (in this region of $x$), and that there is not much hope to observe
the instanton-induced contributions to structure
functions experimentally.
However, instantons are likely to produce events with
a very specific structure of the final state, and such
peculiarities may be subject to experimental search.
A detailed discussion of the hadron distribution in the final state
in the instanton-induced events goes beyond the tasks of this letter.
On general grounds one may expect a resonance-like production
of a fireball of quark and gluon minijects in a narrow region
of Bjorken $x$ of order 0.25 -- 0.40 (for photon-gluon scattering), and
$Q^2$ of order $10^2-10^3 GeV^2$. Such events may be a direct analog
of baryon-number violating processes induced by instantons in the
electroweak theory \cite{ring90}. However, it is not clear whether
in the theory with quark confinement like QCD the structure of the
distributions in the final state can be obtained from the
saddle-point evaluation of functional integrals in the Euclidian
space.

Continuation of the calculation of the
instanton-induced contributions to lower values of Bjorken x
is a difficult problem.
Main open question is the behavior of
the exponential suppression factor. The situation may be quite
different in QCD compared to the electroweak theory, in which
case the unitarization corrections due to multiinstantons
\cite{zakh91} most likely stop the exponential growth of the
instanton-induced cross sections at about one half of the
original semiclassical suppression. The reason is that owing to
the IR divergences
instantons do not produce point-like vertices in QCD,
and it is the external virtuality only, which makes possible the
separation of well-defined finite contributions.
 The multiinstanton
contributions of Maggiore-Shifman type \cite{magg91}
are absent in QCD,
since the virtuality of the photon is able  to cut the
integrations over the size of the first and the last instantons
in the chain only. A more detailed discussion will be given in
\cite{bal94}.  Further theoretical analysis of
high-energy behavior of the instanton-induced cross sections
would be  extremely welcome
and could well jeopardize application of perturbative methods to
the study of
deep inelastic scattering at low values of $x$.


\bigskip
{\bf\large 5.}\hspace{0.5cm}
V.B. acknowledges useful discussions with  M. Beneke, M.A. Shifman
and V.I. Zakharov.
The work by I.B. was supported by the US Department of Energy
under the grant DE-FG02-90ER-40577.

\newpage

\section*{Captions}
\begin{description}
    \item [Fig. 1] The contribution of the instanton-antiinstanton pair
                   to the cross section
                   of hard gluon-gluon scattering (a),
                   structure function of a gluon (b) and of a quark
                   (c,d). Wavy lines are (nonperturbative) gluons.
                    Solid lines are
                   quark zero modes in the case that they are
                    ending at the instanton (antiinstanton),
                    and quark propagators at the $\bar I I$ background
                    otherwise.
    \item [Fig. 2] The non-perturbative scale in deep inelastic
                   scattering  (instanton size $\rho_\ast^{-1}$),
                  corresponding to the solution of saddle-point
                  equations in (\ref{saddleeq}) as a
                  function of $Q$ and for $S(\xi_\ast)\sim 0.5-0.6$
                  ($\xi_\ast\sim 3-4$).
    \item [Fig. 3] Nonperturbative contribution to the structure function
                  $F_1(x,Q^2)$ of a real gluon (\ref{answer})
                  as a function of $x$ for
                   different values of $Q$ (solid curves).
                   The leading perturbative contribution (\ref{box})
                   is shown for comparison by dots. The dashed curves
                   show lines with the constant effective value of
                   the action on the $\bar I I$ configuration.
\end{description}

\clearpage

\begin{figure}
    \begin{center}
        \begin{picture}(110,140)
        \end{picture}
    \end{center}
    \caption[xxx]{
                  The contribution of the instanton-antiinstanton pair
                   to the cross section
                   of hard gluon-gluon scattering (a),
                   structure function of a gluon (b) and of a quark
                   (c,d). Wavy lines are (nonperturbative) gluons.
                    Solid lines are
                   quark zero modes in the case that they are
                    ending at the instanton (antiinstanton),
                    and quark propagators at the $\bar I I$ background
                    otherwise.
                    }
   \label{pic.a}
\end{figure}
\begin{figure}
    \begin{center}
          \begin{picture}(120,110)
          \end{picture}
    \end{center}
    \caption[xxx]{
                  The non-perturbative scale in deep inelastic
                   scattering  (instanton size $\rho_\ast^{-1}$),
                  corresponding to the solution of saddle-point
                  equations in (\ref{saddleeq}) as a
                  function of $Q$ and for $S(\xi_\ast)\sim 0.5-0.6$
                  ($\xi_\ast\sim 3-4$).
 }
   \label{pic.b}
\end{figure}
\begin{figure}
    \begin{center}
\begin{picture}(100,175)
\end{picture}
    \end{center}
\caption[xxx]{
                  Nonperturbative contribution to the structure function
                  $F_1(x,Q^2)$ of a real gluon (\ref{answer})
                  as a function of $x$ for
                   different values of $Q$ (solid curves).
                   The leading perturbative contribution (\ref{box})
                   is shown for comparison by dots. The dashed curves
                   show lines with the constant effective value of
                   the action on the $\bar I I$ configuration.
 }
   \label{pic.c}
\end{figure}
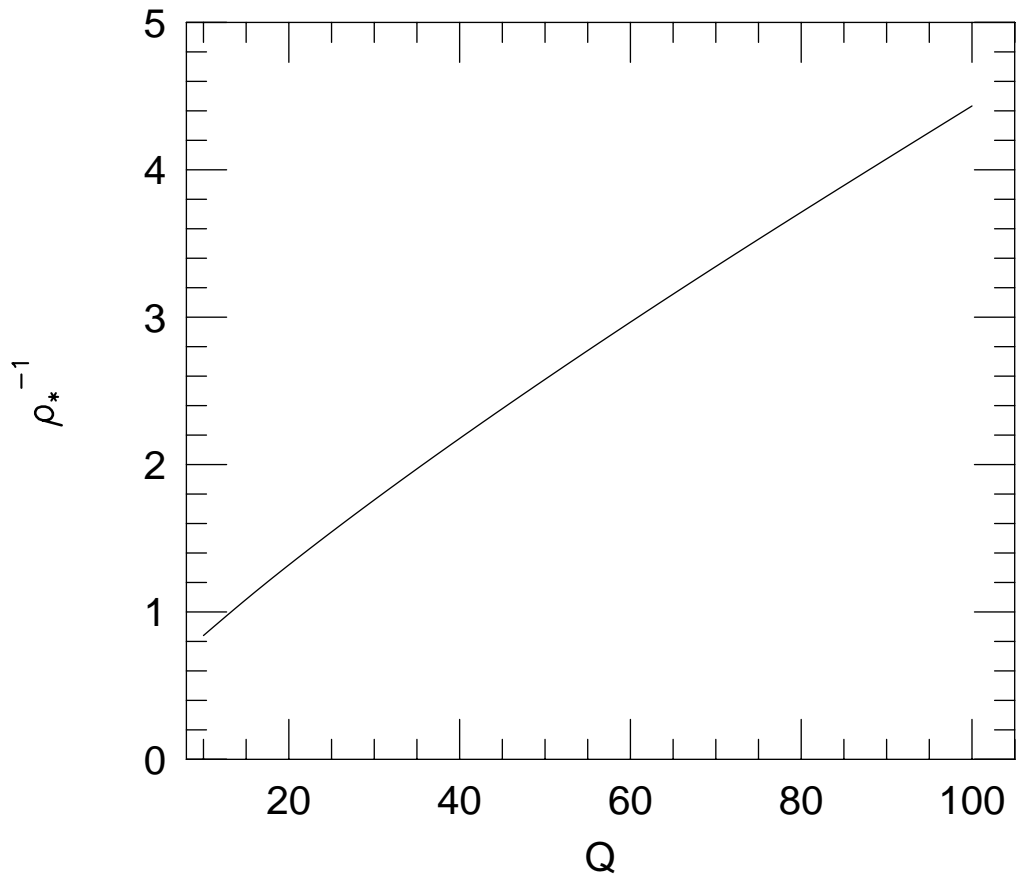
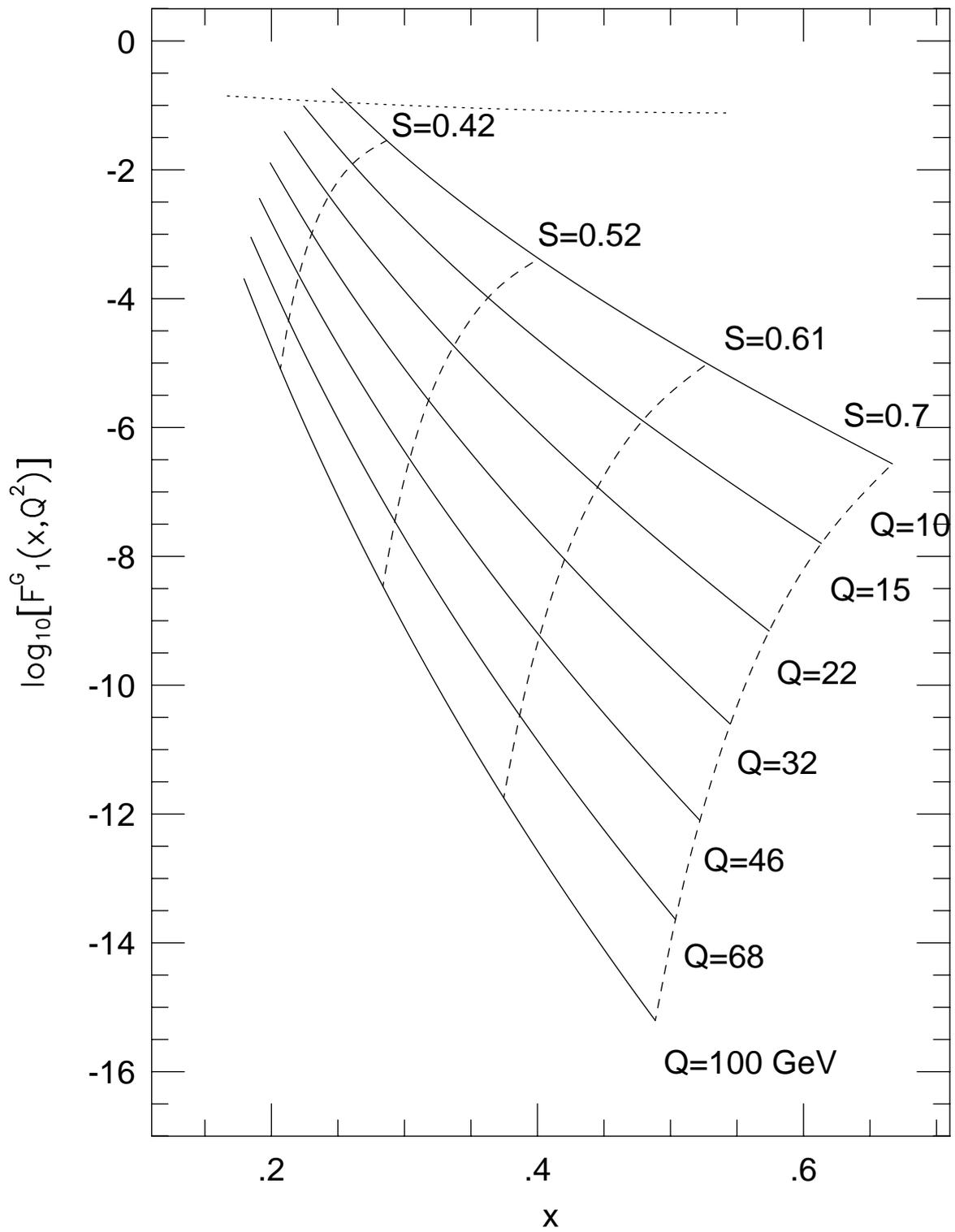
\end{document}